\documentclass{article}
\usepackage{bm}
\usepackage{amsmath}
\usepackage{graphicx}
\usepackage{amsfonts}
\usepackage{setspace}
\usepackage{lineno}

\title{Mapping ammonia emission plumes using shortwave infrared imaging spectroscopy}
\author{
    Nicholas Balasus\textsuperscript{1},
    Daniel H.~Cusworth\textsuperscript{1},
    Jinsol Kim\textsuperscript{1},\\
    Daniel J.~Varon\textsuperscript{2},
    Charles E.~Miller\textsuperscript{3},
    and Riley M.~Duren\textsuperscript{1}\\[1ex]
    \footnotesize\textsuperscript{1}Carbon Mapper, Pasadena, CA, USA\\[-0.75ex]
    \footnotesize\textsuperscript{2}Massachusetts Institute of Technology, Cambridge, MA, USA\\[-0.75ex]
    \footnotesize\textsuperscript{3}Jet Propulsion Laboratory, California Institute of Technology, Pasadena, CA, USA
}
\date{\today}

\begin{document}

\maketitle


\begin{abstract}
Atmospheric ammonia emissions are harmful to ecosystems and human health. These emissions have traditionally been monitored using thermal infrared spectrometers, though such techniques are limited by thermal contrast requirements, the coarse spatial resolution of existing satellite sensors, and low measurement frequency of higher-resolution aerial surveys. Here, we show that ammonia emissions can be quantified using shortwave infrared imaging spectroscopy, circumventing these challenges by using reflected sunlight instead of thermal emission for signal and by enabling a large class of existing and future imaging spectrometers to enter the ammonia observing system. As a proof of concept for this newly discovered capability, we use Tanager-1 satellite data to quantify emissions from industrial point sources of ammonia in Pakistan and Uzbekistan.
\end{abstract}

\section*{Introduction}
Atmospheric ammonia pollutes the environment, both by introducing reactive nitrogen into sensitive ecosystems and by serving as a precursor to hazardous particulate matter \cite{erisman2021}. Currently, agricultural emissions dominate, such as those from synthetic fertilizers or animal waste, while emissions from the energy sector could grow if ammonia is widely adopted as a fuel or hydrogen carrier \cite{bertagni2024}.

Global satellite observations of atmospheric ammonia have identified shortcomings in our understanding of ammonia emissions, including a large underestimate of emissions from industrial processes \cite{vandamme2018}. These observations rely on thermal infrared spectrometers, taking advantage of the strong ammonia absorption features between 10--11 $\mu$m, but their applications to emissions monitoring are limited to large and isolated sources due to the coarse spatial resolution (km-scale) of the observations \cite{dammers2019}. Thermal infrared spectrometers have also been flown on airborne platforms to achieve high spatial resolution, successfully mapping emissions from industrial, agricultural, and natural sources \cite{kuai2019,noppen2023,hasheminassab2025}, though such flights are presently infrequent. Besides these current practical limitations, the requirement for thermal contrast between the surface and atmosphere presents technical challenges, including limiting sensitivity to surface-level emissions and requiring knowledge of plume altitude, the latter of which can greatly complicate emission estimation \cite{noppen2023}.

Here, we show that ammonia absorption features in the shortwave infrared \cite{cacciani2024} allow for emission plumes to be mapped using shortwave infrared imaging spectroscopy, where thermal contrast is not required and high-resolution observations from airborne and spaceborne platforms are more common. As a proof of concept, we use observations from the Tanager-1 satellite to retrieve atmospheric ammonia column enhancements at 30 m spatial resolution, finding large emissions from industrial sources in Pakistan and Uzbekistan. The ability to quantify ammonia emissions using shortwave infrared imaging spectroscopy has the potential to significantly expand the ammonia observing system.

\section*{Methods}
We use calibrated radiance data from the spaceborne Tanager-1 pushbroom imaging spectrometer which has a spectral range covering 400--2500 nm with 5 nm spectral sampling \cite{duren2025}. We use the matched filter algorithm to estimate the path length ammonia concentration enhancements $\hat{\alpha}$ from the at-sensor radiance spectrum $\bm{x}\in\mathbb{R}^n$ for each ground pixel. Here, $n$ is the number of spectral bands, which we limit to the bands between 1400--2500 nm.
\begin{equation}
    \hat{\alpha} = \frac{(\bm{x}-\bm{\mu})^T\bm{\Sigma}^{-1}\bm{t}}{\bm{t}^T\bm{\Sigma}^{-1}\bm{t}}
\end{equation}
We take $\bm{\mu}$ to be the mean radiance spectrum for all pixels observed by the same detector in the along-track direction of the sensor and $\bm{\Sigma}\in \mathbb{R}^{n\times n}$ to be the covariance of the same sample. The target signature $\bm{t}$ is the product of $\bm{\mu}$ and the unit absorption spectrum $\bm{s}$, where $\bm{s}$ describes the relative change in the at-sensor radiance per unit change in ammonia path length enhancement $\alpha$.

To calculate $\bm{s}$, we model the at-sensor radiance spectrum $\bm{I}$ using a non-scattering radiative transfer model \cite{kuhlmann2025} with the U.S. standard atmosphere \cite{anderson1986} with methane and carbon dioxide profiles scaled based on surface concentrations of 1.8 and 420 ppm respectively. Absorption coefficients are from HITRAN2024 using Voigt line shapes \cite{gordon2026}. The high-resolution (0.01 cm$^{-1}$) spectrum $\bm{I}$ is convolved to the instrument-resolution spectrum $\bm{L}$. We then calculate $\bm{s}$ as:
\begin{equation}
    \bm{s} = \frac{\ln \bm{L}_{\text{enh}} - \ln \bm{L}_{\text{bkg}}}{\alpha}
\end{equation}
where $\bm{L}_{\text{bkg}}$ is the modeled at-sensor radiance for the standard atmosphere and $\bm{L}_{\text{enh}}$ is the same but for a surface enhancement of ammonia corresponding to $\alpha$ = 2000 ppm$\cdot$m \cite{knapp2023}. Figure \ref{fig1} shows unit absorption spectra $\bm{s}$ for ammonia as well as methane, carbon dioxide, and water.

We convert the path length enhancements $\hat{\alpha}$ (ppm$\cdot$m) to vertical column mass enhancements $\Delta\Omega$ (kg m$^{-2}$) using the viewing geometry and U.S. standard atmosphere surface conditions. We delineate the plumes using thresholding, a median filter, and dilation, and calculate an ammonia emission rate $Q$ (kg h$^{-1}$) using the integrated mass enhancement method \cite{varon2018}.
\begin{equation}
    Q = \frac{U}{L}\sum_i\Delta\Omega_iA_i
\end{equation}
Here, $U$ (m h$^{-1}$) is the ECMWF IFS 10 m wind, $L$ (m) is the maximum distance between boundary points of the convex hull of the plume mask, and $A_i$ (m$^2$) is the area of each pixel $i$ within the delineated plume \cite{duren2025}. By using this approach, we assume that the lifetime to chemical and deposition losses are sufficiently long relative to transport that they can be ignored.

\section*{Results and Discussion}
Figure \ref{fig2} shows ammonia emission plumes from two industrial fertilizer production plants in Pakistan and Uzbekistan. We infer an emission rate of 837 kg h$^{-1}$ from the fertilizer production plant in Pakistan. Using data from 2013--2017, Dammers et al. \cite{dammers2019} estimated emission rates ranging from 3212--8035 kg h$^{-1}$ for this site depending on the thermal infrared spectrometer used. In Uzbekistan, we quantify two nearby sources with emission rates of 781 and 908 kg h$^{-1}$ for a total emission rate of 1689 kg h$^{-1}$. Using data from 2008--2016, Van Damme et al. \cite{vandamme2018} estimated an emission rate of 2714 kg h$^{-1}$ for the same site. The emission estimates from the coarse resolution thermal infrared sounders that we compare our estimates against \cite{vandamme2018,dammers2019} are sensitive to the ammonia lifetime assumed or fit and may also be sensitive to other nearby sources. Continued sampling by Tanager-1 would allow for more rigorous quantification assessment.

\begin{figure}
    \centering
    \includegraphics[width=0.9\textwidth]{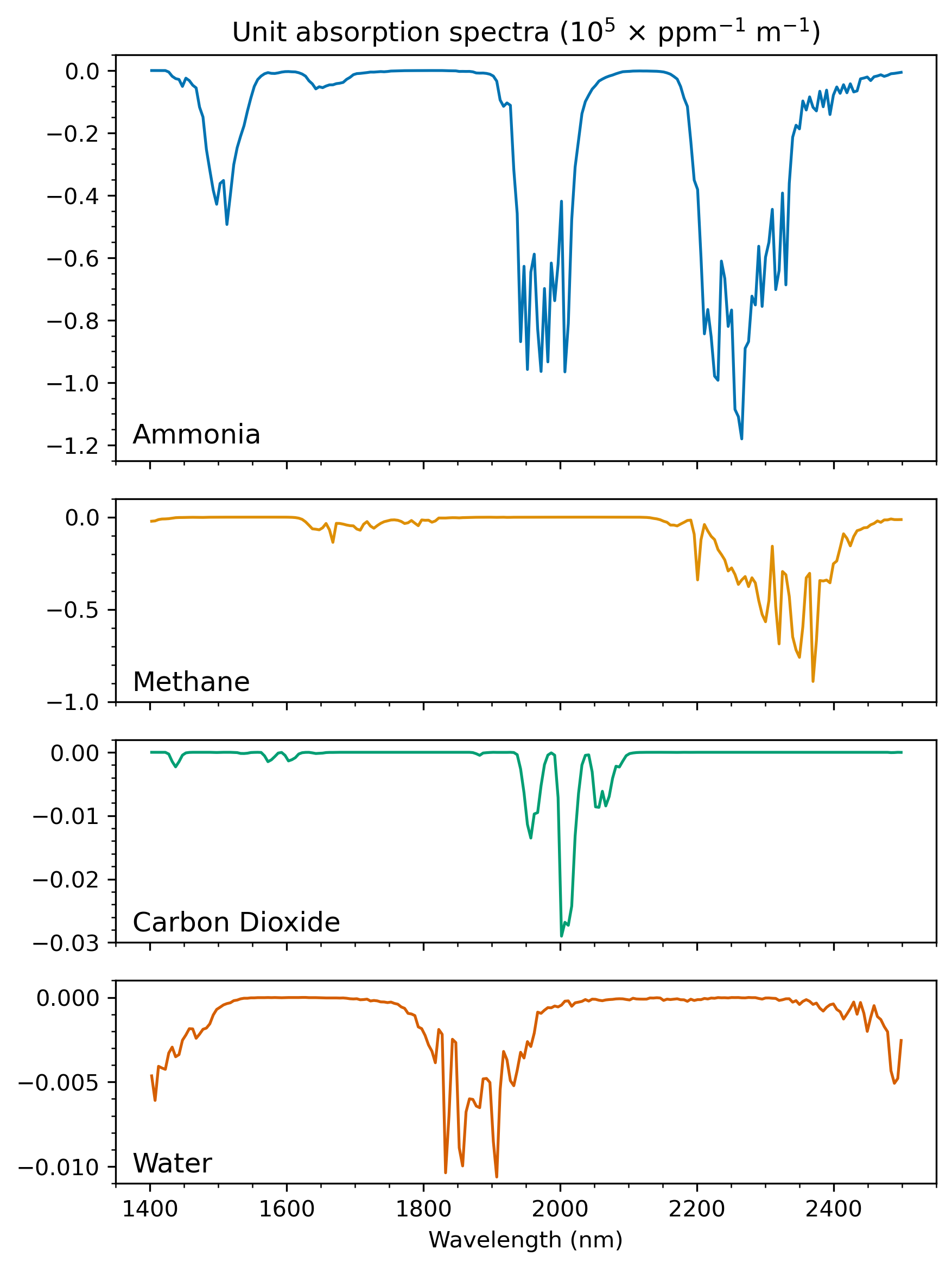} 
    \caption{Unit absorption spectra for ammonia, methane, carbon dioxide, and water for 1400--2500 nm. The spectra depict the relative change in the at-sensor radiance per unit change in a path-length enhancement of the given gas for the Tanager-1 sensor. Values are scaled by $10^5$ for visualization.}
    \label{fig1}
\end{figure}

\begin{figure}
    \centering
    \includegraphics[width=0.9\textwidth]{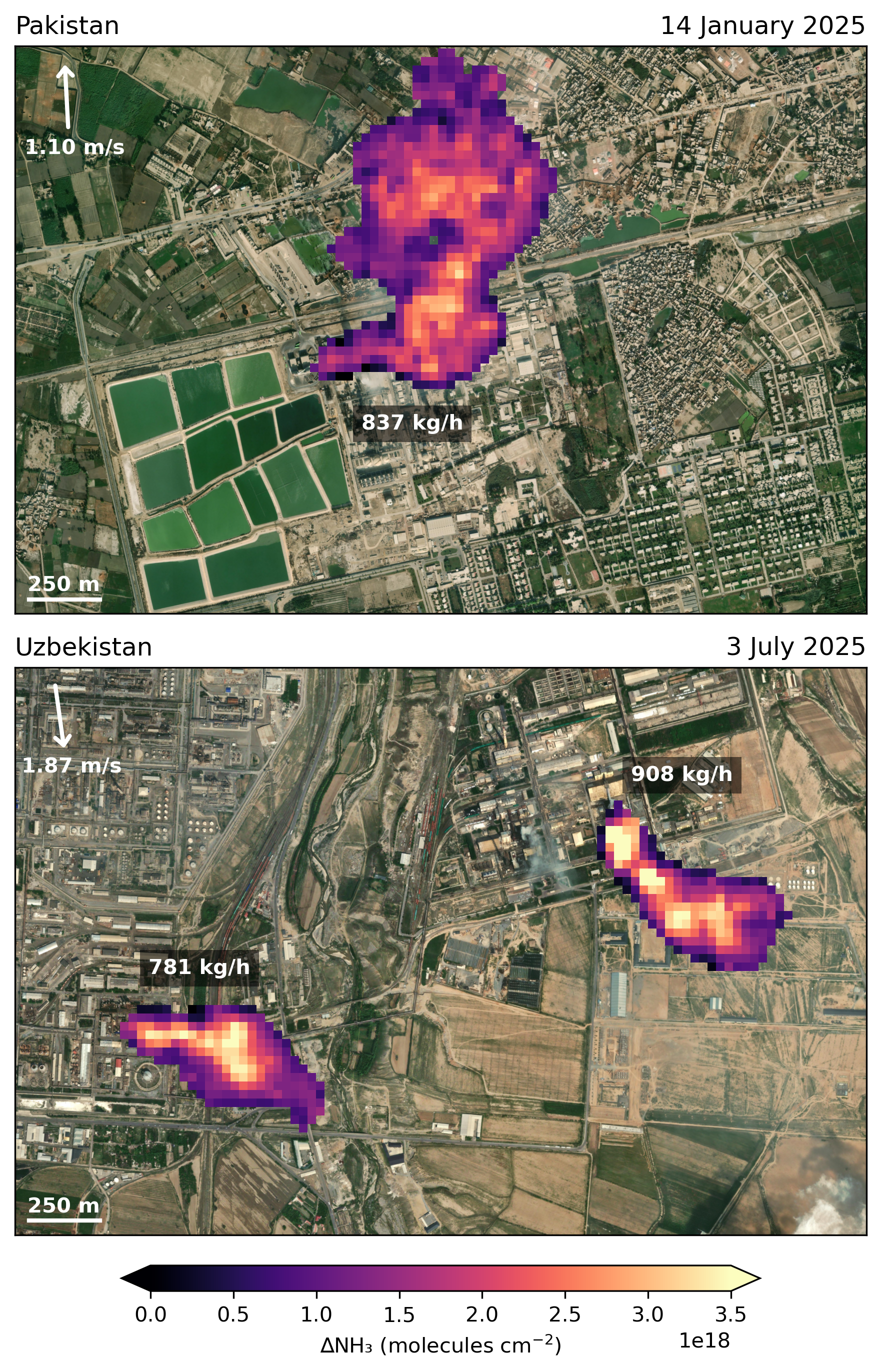} 
    \caption{Ammonia emissions from industrial sources in Pakistan and Uzbekistan mapped using shortwave infrared radiance data from Tanager-1. The winds and the inferred emission rates are inset and ammonia vertical column enhancements are shown. The background imagery is from Esri World Imagery.}
    \label{fig2}
\end{figure}

Our ammonia matched filter retrieval uses radiance data from 1400--2500 nm. To confirm that the ammonia plumes that we have identified are not artifacts due to absorption from other gases shown in Figure \ref{fig1} with absorption features in this range,  we conduct retrievals over the 1425--1575, 1850--2100, and 2175--2300 nm windows. In all cases, the ammonia plumes remain distinct.

This study serves as a proof of concept that ammonia emission plumes can be mapped using shortwave infrared imaging spectroscopy, circumventing technical and practical challenges from the current ammonia observing system focused on the thermal infrared spectral range. There is a class of existing airborne and spaceborne sensors (e.g., AVIRIS, EMIT, PRISMA, EnMAP) that cover the ammonia shortwave infrared absorption features, albeit with variable spectral and spatial resolution and coverage \cite{jacob2022}. Next-generation instruments with high spectral resolution in the 2200--2400 nm range aimed at methane absorption features could offer heightened sensitivity to ammonia.

\clearpage

\clearpage

\section*{Acknowledgments}
The Carbon Mapper team acknowledges the generous support of its philanthropic donors, including the High Tide Foundation, Grantham Foundation for the Protection of the Environment, Bloomberg Philanthropies, and other contributors, that supported this research. Part of this work was carried out at the Jet Propulsion Laboratory, California Institute of Technology, under a contract with the National Aeronatuics and Space Administration.

{
\footnotesize
\bibliographystyle{unsrt}
\bibliography{references}
}

\end{document}